\documentclass[tradiabstract]{aa} 

\usepackage{graphicx,txfonts}
\usepackage{natbib}

\newcommand{\simgreat} {\mathbin{\lower 3pt\hbox{$\rlap{\raise
5pt\hbox{$\char'076$}}\mathchar"7218$}}}

\newcommand{\simless}{\mathbin{\lower 3pt\hbox {$\rlap{\raise
5pt\hbox{$\char'074$}}\mathchar"7218$}}}

\begin{document}

\title{The high-quality single-cloud reddening curve sample}

\subtitle{Dark Dust~{\sc {III}}: Scrutinizing extinction curves in the Milky Way.}

\authorrunning{R.~Siebenmorgen et al.}
\titlerunning{Reddening in the diffuse ISM}

\author {R.~Siebenmorgen\inst{1}, 
  J. Smoker\inst{2,3}, J. Kre{\l}owski\inst{4},
  Karl Gordon\inst{5}, and Rolf Chini\inst{6,7,8}}

\institute{{European Southern Observatory, Karl-Schwarzschild-Str. 2,
85748 Garching, Germany {\tt email: Ralf.Siebenmorgen@eso.org}}
\and {European Southern Observatory, Alonso de Cordova 3107,
Vitacura, Santiago, Chile}
\and  {UK Astronomy Technology Centre, Royal Observatory, Blackford Hill, 
Edinburgh EH9 3HJ, UK}
\and {Materials Spectroscopy Laboratory, University of Rzesz{\'o}w,
Pigonia 1 Street, 35-310, Rzesz{\'o}w, Poland}
\and {Space Telescope Science Institute, 3700 San Martin
Drive, Baltimore, MD, 21218, USA}
\and {Ruhr University Bochum, Faculty of Physics and Astronomy, Astronomical Institute (AIRUB), 44780 Bochum, Germany}
\and {Universidad Cat{\'o}lica del Norte, Instituto de Astronom{\'i}a, Avenida Angamos 0610, Antofagasta, Chile}
\and {Nicolaus Copernicus Astronomical Center, Polish Academy of Sciences, Bartycka 18, 00-716 Warszawa, Poland}
}

\date{Received: August 1, 2022/ Accepted: July 1, 2023 }

\abstract{The nature of dust in the diffuse interstellar medium can be
  best investigated by means of reddening curves where only a single
  interstellar cloud lies between the observer and the background
  source. Published reddening curves often suffer from various
  systematic uncertainties.  We merge a sample of 895 reddening curves
  of stars for which both FORS2 polarisation spectra and UVES
  high-resolution spectra are available. The resulting 111 sightlines
  toward OB-type stars have 175 reddening curves. For these stars, we
  derive their spectral type from the UVES high-resolution
  spectroscopy.  To obtain high-quality reddening curves we exclude
  stars with composite spectra in the IUE/FUSE data due to multiple
  stellar systems. Likewise, we omit stars that have uncertain
  spectral type designations or stars with photometric variability.
  We neglect stars that show inconsistent parallaxes when comparing
  DR2 and DR3 from GAIA. Finally, we identify stars that show
  differences in the space and ground-based derived reddening curves
  between $0.28\,\mu$m and the $U$-band or in $R_V$.  In total, we
  find {{53}} stars with one or more reddening curves passing the
  rejection criteria. This provides the highest quality Milky Way
  reddening curve sample available today.  Averaging the curves from
  our high-quality sample, we find $R_V = 3.1 \pm 0.4$, confirming
  previous estimates. A future paper in this series will use the
  current sample of precise reddening curves and combine them with
  polarisation data to study the properties of {\it {dark dust.}}}


\maketitle


\section{Dust in the interstellar medium}

The disc of our Galaxy, as well as of other spiral galaxies, is filled
with the interstellar medium (ISM), consisting of gas, molecules, and
dust grains. The ISM is clumpy \citep{Dickey1990,Meyer1996},
and covers a wide range of temperatures and
densities \citep{McKeeOstriker}; based on the latter parameter one may
distinguish three categories -- diffuse, translucent, and dense
clouds.

Dense clouds are typically star-forming regions, where the high
extinction ($A_V > 5$\,mag) generally impedes optical observations of
embedded stars. Therefore, this component of the ISM is best analysed
at infrared and microwave wavelengths
e.g. \cite{Binder2018}. Translucent clouds, in contrast, offer the
possibility to study the ISM optically via photometric and
spectroscopic observations of stars, shining through the material at
moderate density of $A_V < 3$\,mag \citep{Snow2006}. The disadvantage
here is that in the majority of cases several clouds are present along
a single-sightline -- an effect that increases with stellar distance.
Likewise, there is the diffuse medium which contributes to the
extinction along the sightline \citep{Li2001}. Furthermore,
translucent clouds show striking differences in their chemical
composition and physical parameters as witnessed by varying intensity
ratios of spectral lines or bands.

The study of the ISM has two origins: Firstly, interstellar gas and
dust were discovered by chance during photometric and spectroscopic
observations of stars \citep{Hartmann1904,Heger1922}, revealing that
the precise knowledge of both the interstellar lines and the amount of
dust along the sightline were crucial for the interpretation of
stellar spectra and photometric data. Secondly, \citep{Spitzer1978}
recognized the fundamental role of the ISM in the process of star and
planet formation.

After the detection of interstellar dust nearly 100 years ago
\citep{Trumpler}, it was soon learned to compare the spectral energy
distribution of unreddened stars with those of reddened stars (see
Sect.~\ref{ISreddening.sec}) to quantify the wavelength-dependant
reddening by dust grains -- leading to the famous {\it extinction
  curve}. Its normalisation $R_V = A_V / E({B-V})$ -- the ratio of
total-to-selective extinction -- was further treated like a constant
of nature with a value of $R_ V \sim 3.1$. Therefore, early model fits
of the extinction curve led to fairly simple dust models with a
power-law grain-size distribution of silicate and graphite
grains. Only photometric studies, some decades later, of stars in
dense, star-forming clouds revealed $R_V-$values of up to five,
suggesting substantial grain growths -- either by coagulation and/or
by the formation of mantles \citep{Mathis1981}. Furthermore, PAHs were
detected by IR spectroscopy in the ISM
\citep{Allamandola1989,Bouwman2001}.  Nevertheless, the $R_V-$value
kept its importance and was suggested to be the only parameter that
determines the extinction law, e.g. \citet{Cardelli89}.

In the present work, we readdress the issue of reddening curves by
analyzing and completing published data. The ideal set of data
required to determine a reddening curve would be a spectroscopically
accurately classified star, with data from the UV to IR, at a reliable
GAIA \citep{Prusti} distance and with a single translucent cloud along
the sightline. Unfortunately, such cases are rare, calling into
question many published reddening curves. For example, due to their
large distances, light from OB-type stars typically passes through
several intervening clouds \citep{Welty1994}. Therefore, the observed
extinction curves for different sightlines are usually quite similar
to each other, simulating a similar $R_V-$value. However, they are
ill-defined averages in the sense that measurements of single-cloud
sightlines with similar chemical compositions should preferentially be
used when studying the physical properties of dust or the magnetic
field direction when extracted from the optical polarisation angle.

To improve the sample from which precise reddening curves can be
derived, we selected sightlines where spectra show interstellar atomic 
lines or features of simple radicals, dominated by only one Doppler
component. Such single-cloud sightlines can be interpreted in terms 
of better defined physical conditions. Furthermore, we focus on 
OB-type stars with known trigonometric parallaxes from GAIA, as
this may allow to estimate both column densities and also
local (volume) densities. For future observations, it seems
important to select and scrutinize all cases from the extensive
catalogues by \cite{V04, FM07, G09}.

This paper is the third in a series concerned with dark dust, which is
composed of very large and very cold grains in the ISM. The first
paper \citep{S20} presented initial results derived from distance
unification and the second \citep{S23} presented a dust model for the
general ISM, which was tested against the contemporary set of
observational constraints \citep{Y20, HD21}. The current paper
presents a high-quality sample of reddening curves { {obtained from
    the literature}} as needed for detailed modelling of particular
sightlines. It is laid out as follows: Sect.~\ref{sample.sec} presents
the sample of sightlines, many of them dominated by single
interstellar lines. Sect.~\ref{ISreddening.sec} describes how
interstellar reddening curves are calculated and discusses the impact
of binarity and uncertainties in the spectral type (SpT) on the
derived curves. The method used to ascribe a SpT to our
sample stars and how uncertainties in the SpT of sample and comparion stars
affect the reddening curve is described in Sect. \ref{spclass.sec}.
Systematic issues affecting the quality of the reddening curves {
  in the literature} are discussed in Sect~\ref{scrut.sec}. The
high-quality sample of reddening curves is presented in
Sect.~\ref{MWredd.sec}, followed by a discussion of the systematic
scatter in published reddening curves and $R_V$ of the same star. A
mean Milky Way reddening curve is derived in Sect.~\ref{MW.sec} and
the main findings are summarized in Sect.~\ref{summary.sec}.


\section{The sample}\label{sample.sec}

Whenever sightlines are observed that intersect different components
of the ISM, the relationship between the physical parameters of the 
dust and the observing characteristics provided by the extinction and 
polarisation data is lost. Hence studying variations of dust properties 
in translucient clouds requires a sample of single-cloud sightlines 
for which the wavelength dependence of the reddening and polarisation 
are available \citep{S18}.

\begin{figure} [!htb]
\begin{center}
\includegraphics[width=9.cm,clip=true,trim=4.7cm 6.7cm 3.2cm 7.3cm]{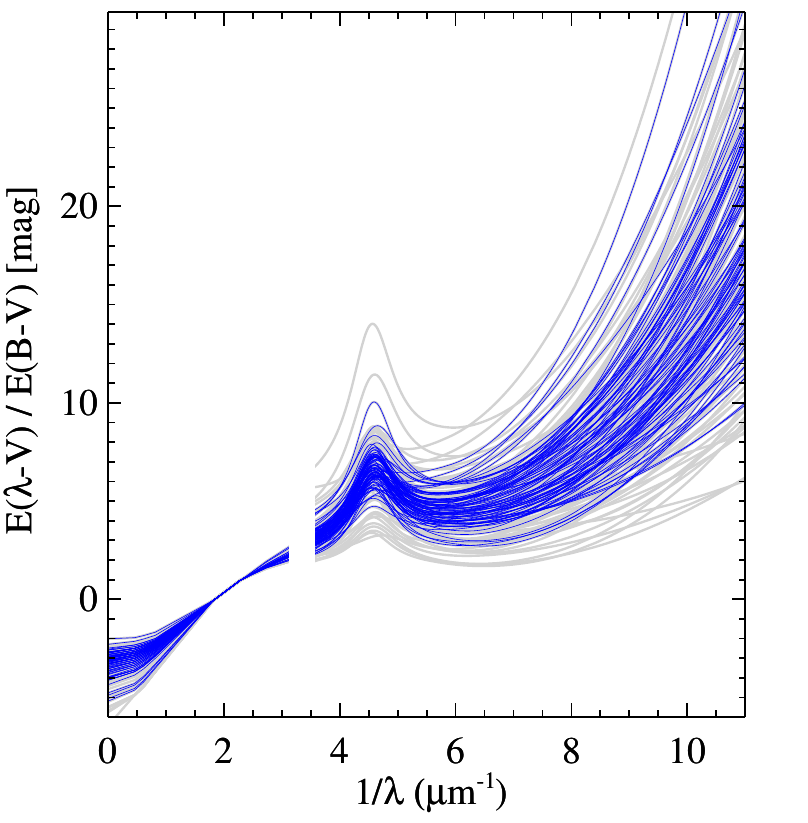}
\end{center}
\caption{Diversity of reddening curves \citep{V04,FM07,G09} of the
  input sample (Sect.~\ref{sample.sec}, gray) and the high-quality
  Milky Way sample (blue, Table~\ref{ok.tab}). \label{redd_all.pdf} }
\end{figure}

Reddening curves have been measured in the near-infrared ($J H K$)
using the Two Micron All Sky Survey (2MASS, \cite{Cutri}), in the
optical ($U B V$) utilizing ground-based facilities
\citep{Mermilliod,V04}, and at shorter wavelengths from space. In
particular, the International Ultraviolet Explorer (IUE) and the Far
Ultraviolet Spectroscopic Explorer (FUSE) provide (far) UV spectra
below 0.3\,$\mu$m down to the Lyman limit. Reddening curves have been
derived from the IUE for 422 stars by \cite{V04}, 351 by \cite {FM07},
including sightlines observed with HST/STIS by \cite{F19}, and for 75
stars with FUSE by \cite{G09}, who adjusted the FUSE spectra to the
larger IUE aperture. In total, 895 reddening curves towards 568 early
types (OB) stars are available from the references described above.

In order to obtain a sample of sightlines with few interstellar components, 
we examined 186 stars mainly observed
with the Ultraviolet and Visual Echelle Spectrograph (UVES;
\citep{UVES1,UVES2} of the ESO Very Large Telescope \citep{S20}. The
term {\it {single-cloud sightline}} was introduced when the 
interstellar line profiles show one dominant Doppler component at a
spectral resolving power ($R$) of $\lambda/\Delta \lambda \sim 75,000$
(full width at half maximum, FWHM $\sim 4$\,km/s) and accounting for
more than half of the observed column density. In total, 65
single-cloud sightlines were detected predominantly by inspecting
the K\,{\sc i} line at 7698~\AA \, \citep{S20}. Single-cloud sightlines
may include two or more fine-structure components in the line profiles
with slightly different radial velocities especially when observed at
an even higher resolution \citep{Welty01, Welty03}. The detection of
65 single-cloud sightlines is a substantial increase as so far only
eight single-cloud sightlines were available for a detailed
analysis. 

Complementary linear polarisation data of 215 stars were taken from
the Large Interstellar Polarisation Survey
\citep{Bagnulo17}. Simultaneous modelling of reddening and
polarisation data provides important constraints of the dust
\citep{S14,S17}. Merging the data of 186 OB stars with UVES spectra,
568 OB stars with 895 reddening curves, and 215 OB stars with
polarisation spectra yields our sample of 111 sightlines. For this
sample, 175 reddening curves are published: 18 by \cite{G09}, 70 by
\cite{FM07}, and 87 by \cite{V04}. They are displayed in
Fig.~\ref{redd_all.pdf}.

\section{Interstellar reddening and extinction}
\label{ISreddening.sec}

Generally, interstellar reddening and extinction curves are derived by
measuring the flux ratio of a reddened and unreddened star with the
same SpT and luminosity class (LC), the so-called standard
pair-method \citep{Stecher}. This method includes uncertainties in the
calibraton of the observed spectra, in the SpT and LC estimates, and
in the need of a close-match in SpT and LC between the reddened and
unredded star, respectively. An alternative to the pair-method is to
use stellar atmosphere models \citep{FM07}. This method avoids the
comparison star but relies on the accuracy of stellar models. In the
current paper we { {examine the accessible databases of reddening
    curves by \cite{V04,FM07} and \cite{G09} for our sample and
apply the following notation:}}

The observed flux of a star is derived from the spectral luminosity
$L(\lambda)$, the distance $D$, and the extinction optical depth
$\tau(\lambda)$, which is due to the absorption and scattering of
photons along the sightline. The observed flux of the reddened star is
given by

\begin{equation} \label{flux.eq}
F(\lambda) = \frac{L({\lambda}) }{ 4 \pi \ D^2} \ e^{-\tau({\lambda})} \ .
\end{equation}

\noindent
We { {follow \cite{K08} and}} denote the flux of the unreddened
($\tau_0=0$) comparison star at distance $D_0$ by $F_0$. The apparent
magnitude is related to the flux through $ m(\lambda) = 2.5 \log _{10}
\Big( w_{\lambda}/F(\lambda) \Big) $, where $w_{\lambda}$ is the zero
point of the photometric system. The difference in magnitudes between
{ {the reddened and the unreddened}} star is $\Delta m(\lambda) = 1.086
\times \Big( \tau(\lambda) + 2 \log _{10}(D/D_0) \Big ) $.

The accuracy of the dust extinction derived by this method depends
critically on the match of both the SpT and LC, and on how well the
distances to both stars are known. Unfortunately, distances to hot,
early-type stars, commonly used to measure interstellar lines, are
subject to large errors \citep{S20}. Hence one relies on relative
measurements of two wavelengths and defines the {\it {color excess}}
{$ E(\lambda - \lambda')=\Delta m(\lambda)-\Delta m(\lambda')$}. The
common notations are e.g. for the $B$ and $V$-band $E(B-V) = (B-V) -
(B-V)_0 $. The {\it {reddening curve}} $E(\lambda)$ is traditionally
represented by a colour excess that is related to the $V$-band and { employs a} 
normalisation to avoid the distance uncertainties, vis:

\begin{equation} \label{eq1}
\begin{split}
E(\lambda) & = \frac {E({\lambda - V})} {E({B-V})} \\
& = \frac {A_{\lambda} - A_V}  {A_B - A_V} \\
& = \frac {\tau_{\lambda} - \tau_V} { \tau_B - \tau_V} \\
\end{split}
\end{equation}

\noindent By definition $E(V) = 0$ and $E(B) = 1$. The extinction in
magnitudes at wavelength $\lambda$ is denoted by $A(\lambda)$. 
An extrapolated estimate of the visual extinction $A_V$ is obtained 
from photometry. This requires measuring $E(B-V)$ and
extrapolating the reddening curve to an infinite wavelength
$E(\infty)$. In practice, $E({\lambda-V})$ is observed at the
longest wavelength which is not contaminated by either dust or any
other emission components of early type stars
\citep{S18b,Deng22}. From this wavelength, e.g. the $K$-band, one
extrapolates to infinite wavelength assuming some a priori shape of
$E(\lambda)$ and hence estimating $E(\infty -V)$. By introducing the
ratio of {\it {total-to-selective extinction}} $R_V = A_V / {E({B-V})}$ 
a simple relation of the reddening to the extinction curve is

\begin{equation} \label{tau2k.eq}
\frac {\tau(\lambda)} {\tau_V} = \frac{E(\lambda)} {R_V} + 1 \,
\end{equation}

\noindent
where obviously $E(\infty) = - {R_V}$. The total-to-selective
extinction of the dust is

\begin{equation} \label{Rv.eq}
\begin{split}
{R_V} & = \frac {\tau_V} { \tau_B - \tau_V} \\
& = \frac {\kappa_V} {\kappa_B - \kappa_V} \, \\
\end{split}
\end{equation}

\noindent where $\kappa = \kappa_{\rm {abs}} + \kappa_{\rm {sca}}$ is
the extinction cross-section which is the sum of the absorption and
scattering cross-section of the dust model.

In the ISM of the Milky Way the total-to-selective extinction scatters
between $2 < {R_V} \simless 4.1$; ${A_V} \sim 3.1 \ E({B-V)}$ is
given as a mean value \citep{V04}. Other approximate formulae using
near IR colours may also be applied, e.g. \cite{Whittet92}
proposed ${R_V \sim 1.1 \ E({V-K}) / E({B-V})}$. We note that a large
$R_V$-value (e.g. $\simgreat 5$) does not necessarily exclude a low
reddening (e.g. $E({B-V}) \simless 0.3$). Indeed, at first sight four 
stars in our sample  show low values of $E({B-V})$ combined with high
$R_V$. However, a more detailed investigation indicates that each of 
these stars has some issue which could impact the derivation of $R_V$
and the reddening curve. In particular, HD~037022 \citep{FM07}
includes in the IUE apertures multiple equally bright objects that
contaminate the observed spectra. The second star HD~037041 shows
photometric instabilities with time variations in the GAIA $G$-band of
0.07\,mag, which is significant, considering that $E({B-V}) = 0.2$
\citep{V04}. Finally, for HD~104705 and HD~164073 varying estimates
of $R_V$ have been derived by different authors, placing doubt on the
``true" value of $R_V$.  As we will show in
Sect.~\ref{systematics.sec}, there is a large systematic error
associated with published $R_V$ estimates of the same star. Hence,
whenever possible, we try to avoid the ${R_V}$ parameter and thus
prefer to discuss reddening instead of extinction curves.


\section{Stellar classification \label{spclass.sec}}

In this section we describe how we determined the spectral classes and
luminosity types of our sample in a uniform manner.

Accurate stellar classification is of utmost importance for deriving
the reddening. A {\sc Simbad} search of the MK classification in our
sample (Sect. \ref{sample.sec}) shows a large spread in the SpT and
the LC. In order to reduce systematics we therefore reclassified the
spectral types of our stars in the MK system using UVES spectra that
were fitted to standard stars. For the standard star spectra we used
the library ``libr18'' by \cite{GC14}. The library includes spectra at
wavelengths between $380 - 462$\,nm. The reduction and analysis of the
UVES spectra are explained in paper~I \citep{S20} and were
complemented with spectra available in the ESO Science Archive
Facility under ESO programme IDs listed in Appendix~\ref{Prog.id}. The
UVES spectra are at a resolving power of $\lambda /\Delta \lambda \sim
75,000$ and high signal-to-noise ($\ge 200$). The different settings
of the UVES spectra were rectified, merged, shifted in wavelength to
match the 410.2\,nm feature, and smoothed by a Gaussian kernel to
equal the spectral resolution of the spectra in the library. SpT and
LC were determined by the best-fitting element of the library to the
spectrum of the target star. The best fit was identified using a
minimum $\chi^2$ condition.

The precision in the classification of O-type stars was estimated by
comparing the \cite{WF90} standards to the SpT derived in the Galactic
O-star survey by \cite{Sota14}. There are 34 O-stars common to both
catalogues. For 32 stars the SpT agrees to better than one subclass and
for two stars the SpT differs by more than that. These are HD~093129
and HD~303308; both are O3 standards by \cite{WF90} whereas
\cite{Sota14} classifies them in agreement to our fitting procedure as
O5 and O4.5, respectively.

Additionally, our SpT and LC estimates were compared with
classifications of early-type standards by \cite{WF90}. These authors
provide 38 O-type and 37 B-type standard stars with different LCs. The
\cite{GC14} library includes 24 spectra for stars between O4 - O9 and
37 spectra for stars between B0 - B9.  Our SpT agrees to better than one 
subclass for 36 stars. A larger spread is only found for two stars;
HD~037043 is classified as an O9V and HD~163758 as an O6.5Ia standard
by \cite{WF90}; in agreement with \cite{Sota14} we assign spectral
types of O7V and O5Ib, respectively.  Most of the 37 B stars are
earlier than B3 with only one B5 and one B8 star.  Our SpT estimates 
match to within better than one subclass for 35 of these stars.  Only
HD~51309 and HD~53138, both classified as B3Iab \cite{WF90}, yield
B5Iab in our fitting procedure.

The goodness of the \cite{GC14} best-fit to the 75 OB standard star
spectra by \cite{WF90} is at minimum $\chi^2 = 0.5 \pm 0.6 < 1.45$ and
the ratio of both spectra varies by $2.4 \pm 0.4 < 3.7$\,(\%). The SpT
estimates of the stars agree to within $\Delta \rm{SpT} = 0.6 \pm
0.3$. \cite{GC14} distinguish luminosity classes I, III, and V. The LC
determination of our procedure agrees with \cite{WF90} to $\Delta
\rm{LC} = 0.5 \pm 0.8$. The accuracy of the classification on the MK
system by our fitting procedure is the same as reached by
\cite{GC14}. These authors report a precision comparable to human
classifiers of $\Delta \rm{SpT} = 0.6$ of a spectral subclass and
$\Delta \rm{LC} = 0.5$ of a luminosity class, respectively.
\cite{Liu19} classified stellar spectra of the LAMOST survey using
\cite{GC14} and confirm the quoted accuracy. A similar precision in
stellar classification is also reached by \cite{Kyritsis22}, again
indicating the accuracy of our classification is on a par with other
works in the literature.

{ {The SpT procedure was also applied to the unreddened comparison
    stars by \cite{Cardelli89} and \cite{G09}. The
    ELODIE\footnote{http://atlas.obs-hp.fr/elodi} and ESO
    archives\footnote{http://archive.eso.org} were inspected for
    available data from high resolution spectrographs ELODIE
    \citep{ELODIE}, ESPRESSO \citep{Espresso}, FEROS \citep{Feros},
    HARPS \citep{Harps}, XSHOOTER \citep{Xshooter}, and UVES.  High
    resolution spectra of 21 standards were found and the various SpT
    estimates agree to previous estimates within one subclass
    (Table~\ref{unreddspt.tab}). }}

\begin{table}[!htb]
\scriptsize   
\begin{center} 
  \caption { {{Spectral types of unreddened comparison stars. \label{unreddspt.tab} }}}
  \begin{tabular}{l l l l l}
    \hline\hline
       1       &    2           &   3        && 4      \\
Standard   & SpT      & SpT$^{L}$				&& Instrument \\
           &     &                              && \\
\hline
HD~047839  &O7~V      	         &O7~V	   &\quad (O7~V)	   	&HARPS \\ 
HD~091824  &O7~V                 &O6~V     &\quad (O7~V)	&FEROS, UVES \\ 
HD~093028  &O8~V       	 	 &O8~V     &\quad (O9~IV)     	&FEROS  \\ 
HD~214680  &O8~V      	 	 &O9~V	   &\quad (O8~V)	&ELODI \\ 
HD~038666  &O8~V/O9~III      	 &O9.5IV   &\quad (O9.5V)   	&HARPS \\ 
HD~210809  &O9~Ia  	 	 &O9~Ib    &\quad (O8~Iab)	&ELODIE \\ 
HD~091983$^V$  &O9~V        	 &O9.5Ib   &\quad (O9~IV)     	&FEROS  \\ 
HD~150898  &B0~Ib 	 	 &B0.5Ia			&&XSHOOTER \\ 
HD~064760  &B0~Ib   	 	 &B0.5Ib   			&&FEROS, HARPS \\ 
HD~036512  &B0~V      	 	 &B0~V      			&&UVES \\ 
HD~046328  &B1~III    	 	 &B1~III    			&&FEROS \\ 
HD~051283  &B1~III    	 	 &B2~III    			&&FEROS       \\ 
HD~055857  &B1~III/B0.5V         &B0.5V    			&&FEROS \\ 
HD~040111  &B1~Ia     	 	 &B1~Ib	  			&&ESPRESSO       \\ 
HD~150168  &B1~Ib     	 	 &B1~Ia	  			&&FEROS \\ 
HD~165024  &B1~Ib     	 	 &B2~Ib	  			&&FEROS \\ 
HD~031726  &B1~V   	 	 &B1~V				&&XSHOOTER \\ 
HD~074273  &B1~V      	 	 &B1.5V    			&&FEROS \\ 
HD~003360  &B2~III		 &B2IV				&&ELODI \\ 
HD~091316  &B2~Ib     	   	 &B1~Iab			&&HARPS \\ 
HD~064802  &B3~V/B2~V        	 &B2~V				&&XSHOOTER \\ 
	   \hline
\end{tabular}
\end{center}
\scriptsize{{\bf {Notes}}: SpT (col.~2) using procedure of
  Sect.~\ref{spclass.sec}, SpT$^{L}$ (col.~3) by \cite{Cardelli89,G09}
  and in paranthesis by \cite{Sota14}. $^V$ Photometric variable star
  (Fig.~\ref{sel_EBV.pdf}).}
\end{table}


\section{Scrutinizing reddening curves \label{scrut.sec}}

Reddening curves offer the possibility of deriving fundamental
characteristics of dust such as particle sizes and abundances.

Systematic issues that affect reddening curves must be minimized for
dust modelling work. To this aim, the trustworthiness of the reddening
curves towards the 111 stars presented in Sect.~\ref{sample.sec} have
been inspected to establish a high-quality reddening curve sample
with predominantly single-cloud sightlines. 

According to Eq.~\ref{ext.eq} reddening curves have a similar shape
making it extremely difficult to exclude less good cases by simple 
inspection (Fig.~\ref{redd_all.pdf}). Even if there exist comparable 
results for the  same star, there might still be errors whenever the
conditions for the derivation of the reddening are not fulfilled. The
quality of the reddening curve depends critically on the precision 
of the SpT and LC estimates, the photometric and spectral stability 
of the reddened and unreddened star, the de-reddening of the 
comparison star, and the photospheric model when applied. In 
general, reddening curves are derived assuming single stellar systems.

In the following subsections, we discuss the systematic effects that
affect the reddening curves of our sample. Uncertain cases and those
that are not qualified for detailed dust modelling are rejected and 
listed in Table~\ref{reject.tab}, while the sample of high-quality 
reddening curves are given in Table~\ref{ok.tab}.

\subsection{Parametrization of reddening curves}
\label{database.sec}

Reddening curves in the UV range between $3.3\,\mu\rm{m}^{-1}
\simless \lambda^{-1} \simless 11\,\mu\rm{m}^{-1}$ are represented by
a spline fit the a Drude profile for the 2175~\,\AA\ extinction
bump, and a polynomial for the UV rise

\begin{equation}
\frac {E({\lambda - V})} {E({B-V}) } = c_1 + c_2 \ x + c_3 \ D(x,\gamma, x_0)
+ c_4 \ F(x) \,
\label{ext.eq}
\end{equation}

\noindent
where $x=\lambda^{-1}$. The Drude profile is given by
\begin{equation}
D(x,\gamma, x_0) = \frac{x^2}{(x^2-x_0^2)^2 + (x \ \gamma)^2} \,
\end{equation}

\noindent
with damping constant $\gamma$ and central wavelength $x_0^{-1}$. 
The non-linear increase of the reddening in the far UV is described by
$F(x)$. \cite{G09} and \cite{V04} applied a form that is given by
\cite{FM90}
\begin{equation}
F(x) = 0.5392 (x - 5.9)^2+ 0.05644 (x - 5.9)^3 \ : \ x \ge
5.9\,\mu\rm{m}^{-1} \,
\end{equation}

\noindent
while \cite{FM07} used
\begin{equation}
F(x) = (x - c_5)^2 \ : \ x \ge c_5 \ .
\end{equation}

\noindent
At longer wavelengths $F(x) = 0$. Following \cite{G09}, we reduce
$c_4$ by 7.5\,\% when data from the IUE alone were used to estimate
the reddening curve, and extrapolate the reddening to $x =
11\,\mu$m$^{-1}$ when necessary. Reddening in spectral regions close
to wind lines at 6.5 and 7.1\,$\mu$m$^{-1}$ and Ly-$\alpha$ at
$8\,\mu$m$^{-1} \leq x \leq 8.45\, \mu$m$^{-1}$, or with apparent
instrumental noise at $ x \simless 3.6\,\mu$m$^{-1}$ is ignored.

Relations between the reddening curve parameters $c_i$ and $R_V$
(Eq.~\ref {ext.eq}) and the dust model parameters are given in Table~5
by \cite{S18}. For example, an uncertainty of 10\% in $c_1$ implies an
uncertainty of 5\% in the the abundance ratio of the large silicate to
carbon particles; a 10\% error in $R_V$ translates into a variation of
the exponent $q$ of the dust size power-law distribution of
10\%. Eventually, a variation of 10\% in $c_4$ results in an
uncertainty in the abundance ratio of very small to large grains of
$\, \simgreat 25$\%.


\subsection{Composite FUSE and IUE spectra \label{composite.sec}}

Reddening curves derived from a composite spectrum that includes the
program star and other bright objects are invalid. Reddening curves of
our sample are flagged when there are multiple objects in the IUE
(separation $\leq 10\arcsec$) or in the FUSE aperture (separation
$\leq 15\arcsec$) that contribute by more than 10\,\% to the flux of
the program star ($\Delta V \simless 2.5$\,mag). The Hipparcos
(ASCC-2.5 \citet{Kharchenko09}) and {\sc Simbad} databases were also
used to detect potential contaminating objects. For 12 stars the label
M is assigned in Table~\ref{reject.tab} indicating that the observed
spectrum is a composite of multiple bright objects in the IUE
aperture. No companions were found in the FUSE aperture.

\subsection{Multiple star systems \label{multi.sec}}

The multiplicity of stellar systems is typically investigated through
imaging and interferometry \citep{Sana14} and/or spectroscopy
\citep{C12}. The latter method is biased towards finding close
companions down to a few AU separation by providing a measure of time
variable radial velocities.  Differences in the line profiles are
still visible in inclined systems for which the brightness difference
between primary and companion may become marginal. Such a high
resolution radial velocity survey with spectra taken at multiple
epochs (2-12) of about 800 OB stars is presented by
\cite{C12}. Companions in that survey are detected down to a
brightness difference of $\Delta V \sim 2$\,mag. This translates to
detectable companions of an O5 star range between O5 - B2, and those
of a B9 star from B9 - A7. The results of the surveys listed above
indicate that nearly 100\,\% of O-type stars have one or more
companions within 1\,mas to 8$''$ separation and at a contrast down to
$\Delta H = 8$\,mag \citep{Sana14}, falling to 80\,\% for early-B and
to 20\,\% for late B-types \citep{C12}. The HIPPARCOS Tycho
photometric catalogue \citep{Kharchenko09} provides various labels
indicating the duplicity or variability status of the
stars. Unfortunately, there are no striking features in the reddening
curves noticeable when inspecting sub-samples with variability/binary
flag set or unset.

\subsection{GAIA \label{gaia.sec}}

The GAIA space observatory launched by { the European Space Agency}
(ESA) in 2013 measures positions, parallaxes, motions, and photometry
of stars with unprecedented precision \citep{Prusti}. The GAIA data
release 2, DR2 \citep{DR2} was based on observations made between July
2014 and May 2016 and was followed by data release 3, DR3 \citep{eDR3,
  DR3} which includes observations until May 2017.

\begin{figure} [h!tb]
\begin{center}
\includegraphics[width=9.cm,clip=true,trim=4.cm 9cm 3cm 9.3cm]{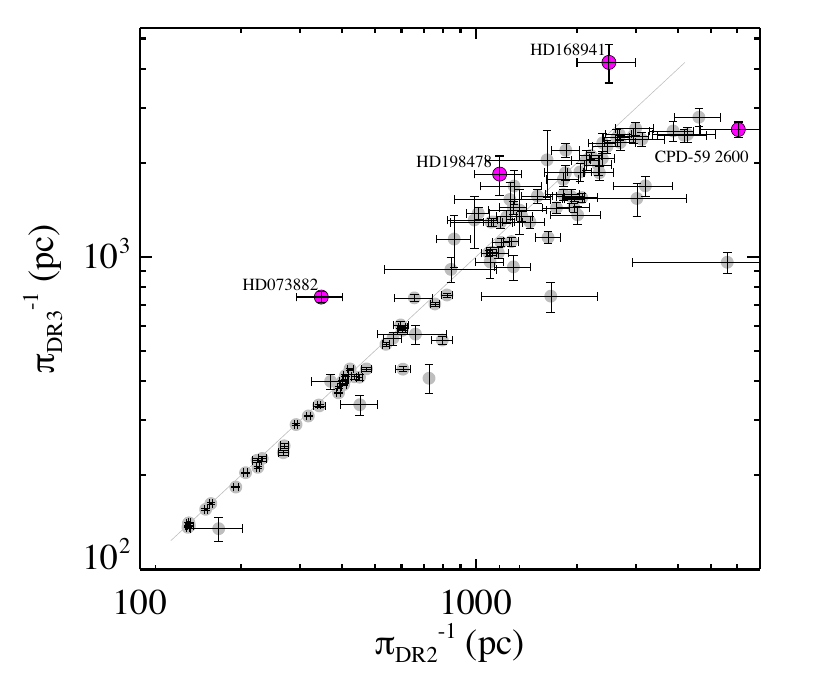}
\caption{\label{GAIApar.pdf} Inverse parallax (pc) of GAIA data
  releases DR3, $\pi^{-1}_{\rm {DR3}}$, versus DR2 , $\pi^{-1}_{\rm
{DR2}}$. Stars labelled in magenta are rejected. The identity of
$\pi^{-1}_{\rm {DR3}}$ is shown as grey line. \label{pl_GAIA_par.pdf}}
\end{center}
\end{figure}

Stars that have inconsistent GAIA parallaxes $\pi_{\rm {DR2}}$ versus
$\pi_{\rm {DR3}}$ are suspicious. Their parallax measurements at
higher than $3 \sigma$ confidence remain unconfirmed either due to
instrumental artefacts, which we doubt, bright companions or stellar
activity. In our sample, there are 102 stars with parallax
measurements in DR2 at a typical S/N ratio of 13; in DR3 there are 106
of our stars with a typical S/N ratio of 25. From these stars, 95 have
a ratio $\pi_{\rm {DR2}}/ \pi_{\rm {DR3}} $ higher than $3\sigma$
confidence with a mean ratio of $0.94 \pm 0.16$. The 6\,\% deviation
from unity is driven by unsecured DR2 measurements at large
distances. By considering a sub-sample with $\pi_{\rm {DR2}}^{-1}
\simless 2$\,kpc there are 72 stars with a mean ratio of $\pi_{\rm
  {DR2}}/ \pi_{\rm {DR3}} \sim 0.98 \pm 0.14$ which is consistent with
being unity. Four stars are identified by $3\sigma$ clipping as
outliers and are indicated with the flag $\pi$ in
Table~\ref{reject.tab}. The inverse parallax (pc) of $\pi^{-1}_{\rm
  {DR3}}$ versus $\pi^{-1}_{\rm {DR2}}$ is shown with outliers marked
in magenta in Fig.~\ref{GAIApar.pdf}. Note the deviation from the
identity curve at $\pi^{-1}_{\rm {DR2}} \simgreat 2$\,kpc.

The photometric stability of the stars of our sample was verified by
comparing both GAIA data releases. In Fig.~\ref{pl_GAIA_par.pdf} the
differences in $G$-band ($330 - 1050$\,nm) photometry between DR3 and
DR2, $\Delta G = G_{\rm {DR3}} - G_{\rm {DR2}}$, is shown for our
sample of 111 stars. In that sample, the mean and $1\sigma$ scatter of
$\Delta G = 14 \pm 11$\,mmag. Nine stars show variability in the
photometry with $\Delta G$ outside of the range $14 \pm
33$\,mmag. They are marked in Fig.~\ref{pl_GAIA_par.pdf} in magenta
outside the $\pm 3 \sigma$ curves shown as dashed lines. Their
reddening curves are classified as uncertain.  Six of these sightlines
were not yet rejected and they are labelled $\Delta G$ in
Table~\ref{reject.tab}. 


\begin{figure} [h!tb]
\includegraphics[width=9.5cm,clip=true,trim=3.cm 6cm 2.5cm 7cm]{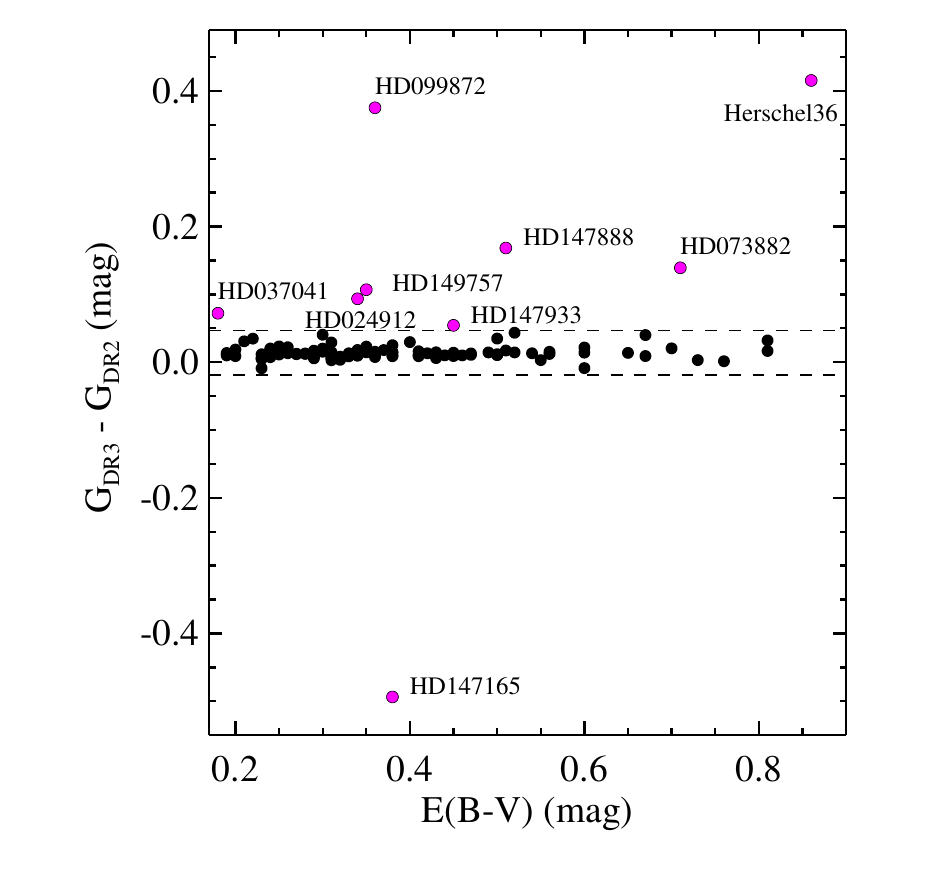}
\caption{\label{selGAIA.pdf} The differences in GAIA $G$-band
  photometry between data release DR3 and DR2, $G_{\rm {DR3}} -
  G_{\rm {DR2}}$, as a function of reddening $E(B-V)$. Stars labelled
  in magenta outside the dashed lines are rejected. }
\end{figure}


\noindent 
We note that seven stars from our sample would be rejected due to
colour variability in $B - G$, whereas stars outside of the range
$(B-G)_{\rm {DR3}} - (B-G)_{\rm {DR2}}$ of $31\pm 66$\,mmag would be
excluded using 3$\sigma$ clipping. These stars were already rejected
using the previous criteria.


\subsection{Photometric variability \label{var.sec}}

Stellar variability will impact the results for reddening. 
Early-type stars may vary due to multiplicity or due to winds. 
About 50\% of OB stars reveal extra-photospheric infrared excess 
emission that is likely caused by winds \citep{S18b,Deng22}, 
and strikingly, as discussed in Sect.~\ref{multi.sec}, the great 
majority of O and early-type B stars form close binary systems.

The reddening curves that we obtained from the literature made 
use of the Johnson $UBV$ system \citep{Hiltner56, Nicolet78}. 
Ground-based (GB) photometry of our sample is given by
\cite{V04} and refers to observations between $1950 - 2000$. 
Stellar photometry from the Hipparcos satellite between $1989 -
1993$ is also available. \cite{Kharchenko09} merged Hipparcos, Tycho,
PPM, and CMC11 observations of 2.5 million stars and transformed $V$
and $B$ magnitudes to the Johnson system. This resulted in a
colour-dependent correction of $20 - 40$\,mmag, and a typical error
below $10$\,mmag of the ASCC-2.5 catalogue \citep{Kharchenko09}.

\begin{figure} [h!tb]
\includegraphics[width=9.5cm,clip=true,trim=3.cm 6cm 2.5cm 7cm]{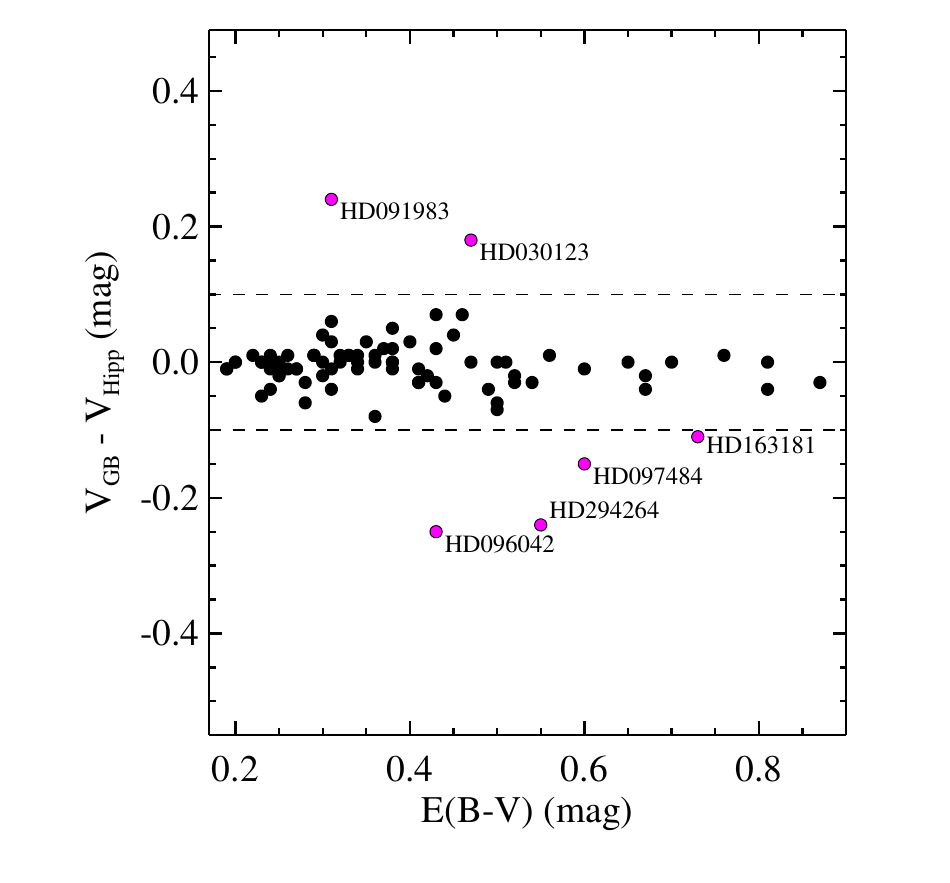}
\caption{\label{sel_EBV.pdf} The differences in $V$-band photometry
  between ground-based \citep{V04} and Hipparcos
  \citep{Kharchenko09} $V_{\rm {GB}} - V_{\rm {Hip}}$ as a
  function of reddening $E(B-V)$. Stars labelled in magenta outside the
  dashed lines are rejected.}
\end{figure}

The photometric stability of the stars was verified by comparing
ground-based $V_{\rm {GB}}$ \cite{V04} and Hipparcos $V_{\rm {Hip}}$
\cite{Kharchenko09} photometry (Fig. \ref{sel_EBV.pdf}). There are 73
stars for which photometry in both catalogues is available and which
are not rejected by the previous confidence criteria
(Sect.~\ref{composite.sec}~-~\ref{gaia.sec}).  In that sample the mean
and $1\sigma$ scatter of $\Delta V = V_{\rm {GB}} - V_{\rm {Hip}} = 5
\pm 31$\,mmag. The small offset is within the photometric error. Seven
stars are identified by $3\sigma$ clipping as outliers showing
significant variability in the $V$-band. Their red-


\noindent dening curves are classified as uncertain and labelled
$\Delta V$ in Table~\ref{reject.tab}. These stars are marked in
magenta in Fig.~\ref{sel_EBV.pdf} and are outside the $\pm 3 \sigma$
variation shown as dashed lines. HD~037023 was identified as a
composite star in Sect.~\ref{composite.sec} and fails with $\Delta V =
1.6$\,mag.  { {The photometric variable star HD~091983 was used by
    \cite{G09} to derive the reddening curves towards HD~122879
    (Table~\ref{unreddspt.tab}) and HD~168941.}}

The same procedure of outlier rejection was repeated by comparing the
$B$-band photometry and the $(B - V)$ colour provided from the ground
by \cite{V04} and by Hipparcos \citep{Kharchenko09}. The mean and
$1\sigma$ scatter in the colours is $\Delta_{BV} = (B-V)_{\rm {GB}} -
(B-V)_{\rm {Hip}} = -28 \pm 28$\,mmag. The two stars HD~147701 and
HD~169454 are identified by $3\sigma$ clipping as outliers. Their
reddening curves are classified as uncertain and labelled
$\Delta_{BV}$ in Table~\ref{reject.tab}. No additional stars were
rejected due to photometric variability in the $B$-band.

\subsection{Unfeasible stellar classification}

O- and B-type stars are often fast rotators; the peak in the rotational 
velocity probability distribution for B-type stars is around 300\,km\,s$^{-1}$  
\citep{Dufton19}. As a consequence their SpT and LC
determination is highly uncertain because: (1) most useful diagnostic lines
such as Mg\,~{\sc{ii}} or He\,~{\sc{i}} are blended and (2) unless the
spectra have a very high S/N ratio, many stellar absorption lines
merge into the continuum. Several of the stars have also a bright
companion making the stellar classifcation unfeasible. For nine stars
the stellar classification procedure of Sect.~\ref{spclass.sec} is
uncertain at $\chi^2 > 2$. These sightlines are rejected for the
high-quality reddening curve determination with label $\chi^2$ in
Table~\ref{reject.tab}.

\subsection{Inaccurate stellar classification \label{splredd.sec}}

A spectral type and luminosity miss-match of the target or comparison
star {{can give a}} large variation in $E(B-V)$, in the infrared
extinction, differences in position and width of the extinction bump,
and in the far UV rise of the reddening curve. { {As shown by
    \cite{Massa83, MF86, Cardelli92}, photometric and systematic
    errors of the extinction curves scale as $1/E(B-V)$. A
    miss-classification in the LC or SpT of more than one subclass in
    either of the reddend or the unreddened star may introduce large
    ($\sim 20$\,\%) systematic errors in $E(B-V)$ and a significant
    difference in the far UV rise \citep{Cardelli92}.  They also show
    that such a change in the far UV rise expressed in parameter c4
    (Eq.~\ref{ext.eq}) may vary by a factor $\sim 1.5$}}. Whenever the
comparison star is hotter than the reddened star the far UV rise will
be overestimated. In the UV, the IUE and FUSE spectrographs were used
to estimate SpT and LC. Besides the low resolving power of { these}
UV spectrographs, there are major diagnostic lines in the optical but
not in the UV.  \citet{SmithNeubig97} show that the UV spectral
diagnostics indicate often earlier SpT than obtained from optical
spectra. They considered uncertainties of one luminosity class and one
spectral sub-type, which increases to up to two sub-types for mid/late
B stars because of fewer spectral diagnostics in that range.

\cite{G09} list the SpT of the reddened and comparison stars.
\cite{V04} apply comparison stars selected from \citet{Cardelli92},
who provide those for types earlier than B3. For later stars, the
choice of the comparison star was not detailed; {{same holds for
    stars earlier than O7. We note that the stellar atmosphere
    model-based method, as used by \citet{FM07}, does not need to
    apply a comparison star}.}  For some stars we find differences in
the SpT and LC of more than one subclass when derived from
observations in the UV \citep{V04,FM07,G09} and in the optical
(Sect.~\ref{spclass.sec}). This introduces systematic errors in the
reddening curves as mentioned above. This { {affects fifteen}}
  curves by \cite{V04}, two by \cite{G09}, and six by \cite{FM07}. For
  the latter there is an extra difficulty that the temperature of the
  \cite{Lanz07} model atmospheres needs to be related to the MK system
  by means of a stellar temperature scale \citep{Theo91}. In that
  scheme we add an extra uncertainty of 1,000\,K in favour of
  non-rejection.

{ {In 40 out of 54 cases, the SpT of the reddened and the
    comparison stars agree within 1.5 subclasses. These stars are
    indicated below the line in Table~\ref{compspt.tab} and are kept
    in the high-quality sample, whereas the cases above that line show
    larger deviations. Their reddening curves are considerered to be
    of lower quality. }} For example, the spectral type of HD~093843
is uncertain; we find O4~Ib, \cite{Sota14} O5~III, and \cite{V04}
O6~III.  For deriving the reddening curve \cite{V04} used an O7~V
comparison star, which deviates by more than two types; therefore, we
reject the reddening. HD~046660 whose spectral type is also uncertain,
when using the fitting procedure (Sect.~\ref{spclass.sec}), a similar
$\chi^2$ minimum is found for either B0~III or O7~V. We adopt the
latter type as the star displays He\,{\sc ii}. It has been classified
as O9~V by \cite{S20} implying a temperature $\ga 30,000$\,K,{{ in
    agreement with 31,067\,K derived \cite{F19}. However, \cite{FM07}
    derived the reddening with a}} best fitting model atmosphere at
26,138\,K, hence later than B1 \citep{Theo91, PM13}. Additionally,
\cite{V04} derived the reddening with a B1.5~V comparison star. Due to
the discrepancy between these results and our derived spectral type,
we also reject this reddening curve. { {As detailed in
    Table~\ref{compspt.tab}, the reddening curve of HD~168076
    (O4~III/O5~V) was derived by \cite{V04} using an O5~V type star,
    while \cite{G09} used HD~091824 (O7~V) and is therefore
    removed. For HD~167771 the reddening curve by \cite{V04} was
    derived with a comparison star that differs by two subclasses and
    is removed, while the \cite{G09} derived curve with a smaller SpT
    difference is included (Table~\ref{compspt.tab}).  The reddening
    curve towards HD~108927 (B5~V) was derived by \cite{V04} using a
    B3~V comparison star.  Reddening curves derived with comparison
    stars with unconfirmed SpT in Simbad are kept. For example,
    \cite{G09} used as comparison stars HD~051013 (B3V?) for deriving
    the reddening of HD~027778 (B3~V) and HD~062542 (B5~V).  Both
    reddening curves show (Fig.~6) excellent agreement with those
    derived by \cite{F19}, respectively.  }} Further comments on
individual stars are given in Appendix~\ref{comm.ap}. In total, 15
sightlines have uncertain reddening curves because of a SpT mismatch
{{(Table~\ref{compspt.tab})}}. They are flagged in
Table~\ref{reject.tab} by attaching the SpT and LC to that star.

\begin{table}[!htb]
\scriptsize   
\begin{center} 
  \caption { { {Comparison of spectral type between the reddened
        and the unreddened star. \label{compspt.tab} }}}
  \begin{tabular}{l c | l l}
    \hline\hline
       \multicolumn{2}{c|}{Reddened star} & \multicolumn{2}{c}{Unreddened star} \\
\hline 
 O4~Ib         & HD~093843      & O7~V       & HD~047839  \\
 O4~Ib         & HD~153919      & O9.5Ia    & HD~188209  \\
 O7~V          & HD~046660      & B1.5V     & HD~074273  \\
 O8~Ia         & HD~151804      & O9.5Ia    & HD~188209  \\ 
 O8~II 	       & HD~162978      & O9.5IV    & HD~188209  \\
 O8~V          & HD~164816      & B0~V       & HD~097471$^G$    \\
 B2~Ib         & HD~103779      & B0.5Ib    & HD~064760 \\
               & $"$            & B0.5II    & HD~094493$^G$  \\
 B2~Ib         & HD~185859      & B0.5Ib    & HD~064760  \\
 B3~V	       & HD~315033      & B1.5V     & HD~074273  \\
 B3~V          & HD~054306      & B1~V       & HD~031726 \\ 
 B5~Ib         & HD~072648      & B1.5III   & HD~062747  \\
 B5~III	       & HD~203532      & B3~IV      & ?  \\
 B8~III	       & HD~134591      & B4~V       & ? \\ 
 \hline
  O4~III/O5~V  & HD~168076  	& O5~V       & ?	 \\
               &  $"$          & O7~V       & HD~091824$^G$    \\
 O4~V          & HD~046223  	& O5~V       & ? \\
 O4~V          & HD~093205  	& O5~V       & ? \\
 O6~V          & HD~303308  	& O5~V       & ?	 \\
 O7~III        & HD~167771  	& O5~V       & ?	 \\
               & $"$            & O8~V       & HD~093028$^G$   \\
 O7~III        & HD~093222  	& O7~V       & HD~047839       \\
               & $"$            & O8~V       & HD~093028$^G$ \\
 O8~V          & HD~046149  	& O9~V       & HD~214680 (O8~V) \\
 O9~Ia         & HD~152249  	& O9.5Ia    & HD~188209       \\
               & $"$            & O9~Ib      & HD~210809$^G$     \\
 O9~V          & HD~046202  	& O9.5IV    & HD~038666 (O8~V/O9~III) \\
               & $"$            & O8~V       & HD~093028$^G$     \\
 B0~Ia         & HD~122879      & B0~Ib      & HD~204172  \\
               & $"$            &  ?         & HD~091983$^{G}$     \\
 B0~Ib         & HD~047382  	& B0~III     & HD~063922  \\
 B0~Ib         & HD~164402  	& B0~Ib      & HD~204172  \\
 B0~Ib         & HD~167264  	& B0~Ib      & HD~204172  \\
 B0~III        & HD~101008  	& B0~V       & HD~036512  \\
 B0.5~III      & HD~152245  	& B0.5IV    &  ? \\
 B0.5~V        & HD~185418  	& B0.5III   & HD~119159  \\
               & $"$            &  ?         & HD~097471$^G$      \\
 B1~Ia         & HD~152235      & B0.5Ib    & HD~064760  \\
 B1~Ib         & HD~092044      & B1.5III   & HD~062747  \\
 B1~V          & HD~054439      & B1.5V     & HD~074273  \\
 B1~V          & HD~129557      & B1.5III   & HD~062747  \\
 B1~V          & HD~315032      & B1~V       & HD~031726  \\
 B2~Ia         & HD~148379      & B2~Ib      & HD~165024  \\
 B2~III        & HD~110946      & B2~III     & HD~051283 (B1~III) \\
 B2~V          & HD~170740      & B1~V       & HD~031726  \\
 B3~V          & HD~027778      & B3~V?      & HD~051013$^G$     \\
 B3~V          & HD~038023      & B4~IV      & ? \\
 B2~V          & HD~037903      & B1.5V     & HD~074273  \\
               & $"$            & B3~V?      & BD~+52$^{\rm o}$3210 \\
 B2~V          & HD~315023      & B1~V       & HD~031726  \\
 B3~III        & HD~070614      & B4~IV      & ? \\
 B5~V          & HD~062542      & B4~V       & ?	 \\
               & $"$            & B3~V?      & HD~051013  \\
 B7~V          & HD~096675      &  ?         & HD~037525$^G$      \\
\hline
\end{tabular}
\end{center}
\scriptsize{\bf{ Notes:}} { { SpT estimates using the procedure of
    Sect.~\ref{spclass.sec} (Table~\ref{unreddspt.tab}) are added in
    parenthesis when deviating from the SpT used by \cite{V04} or
    \cite{G09}, marked by $^G$.}}
\end{table}


\subsection{Comparison of reddening derived from the IUE and Ground-based observations \label{E028u.sec}}

\begin{figure} [h!tb]
\includegraphics[width=9.5cm,clip=true,trim=3.cm 6cm 2.5cm 7cm]{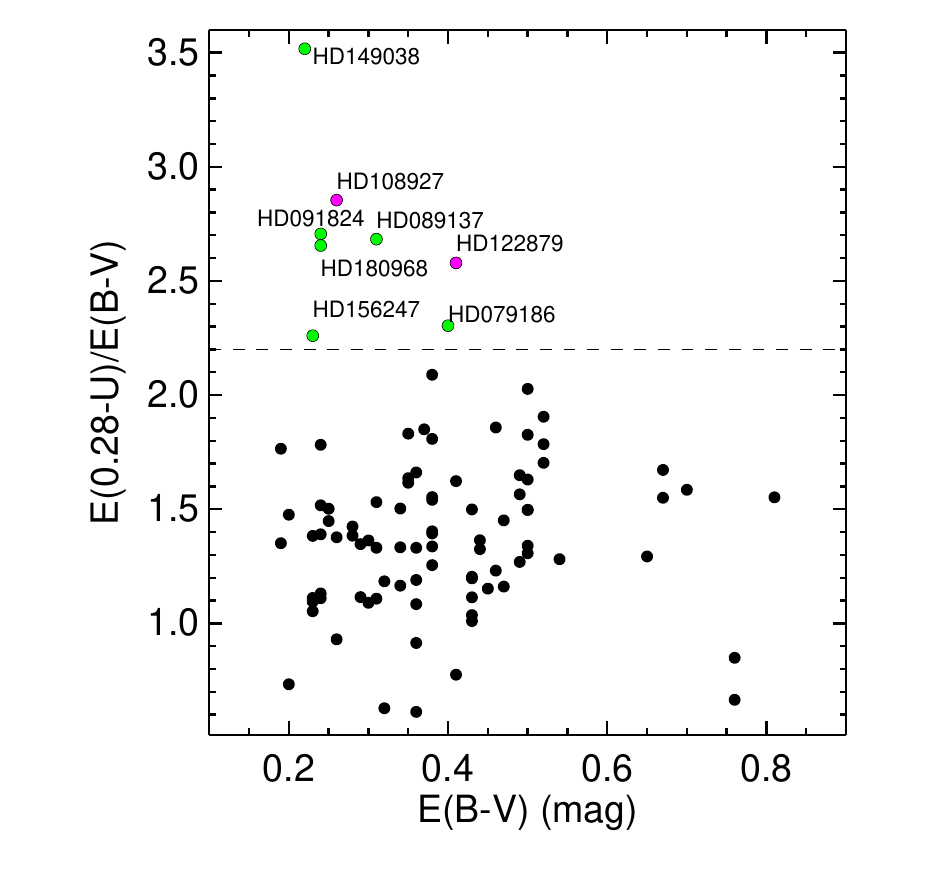}
\caption{\label{pl_Uiue.pdf} The reddening between $0.28\,\mu$m and
  the $U$-band as a function of $E(B-V)$.  {{The reddening curves
      towards the stars above the dashed line were derived by
      \cite{V04}. They are marked in green when included in the
      high-quality sample (Table~\ref{ok.tab}) and in red when
      rejected.}}}
\end{figure}

After applying the confidence criteria, we discovered that some of the
remaining reddening curves display a jump when going from the longest
wavelength of the IUE spacecraft, that is not contaminated by
instrumental features at $0.28 \, \mu$m, to the shortest wavelength
accessible from the ground, the $U$-band. The reddening between
$0.28\,\mu$m and the $U$-band for these 96 curves has a mean and $1
\sigma$ scatter of $E$(0.28$\mu$m - $U$)/$E(B-V) = 1.3 \pm 0.3$~mag
and are shown in Fig.~\ref{pl_Uiue.pdf}. Peculiar reddening curves
that show an offset in $E$(0.28$\mu$m - $U$)/$E(B-V)$ were identified
by means of $3 \sigma$ clipping. { {These include eight reddening
    curves which are derived by \cite{V04}. Two of these, HD~108927
    and HD~122879, were rejected above. For HD~079286, HD~089137, and
    HD~149038 no other reddening curve is available.  For the
    remaining three sightlines -- HD~091824, HD~156247, and HD~180968
    -- only the reddening curves derived by \cite{V04} show this
    striking jump whereas this feature is not present in the reddening
    curves derived by \citet{FM07} for the same stars
    (Fig.~\ref{reddFV.pdf}).  These six reddening curves are marked as
    peculiar in Table~\ref{ok.tab}.}}


\begin{figure*} [h!tb]
 \begin{center}
   \includegraphics[width=18cm,clip=true, trim=4cm 10cm 3.5cm  6.9cm]{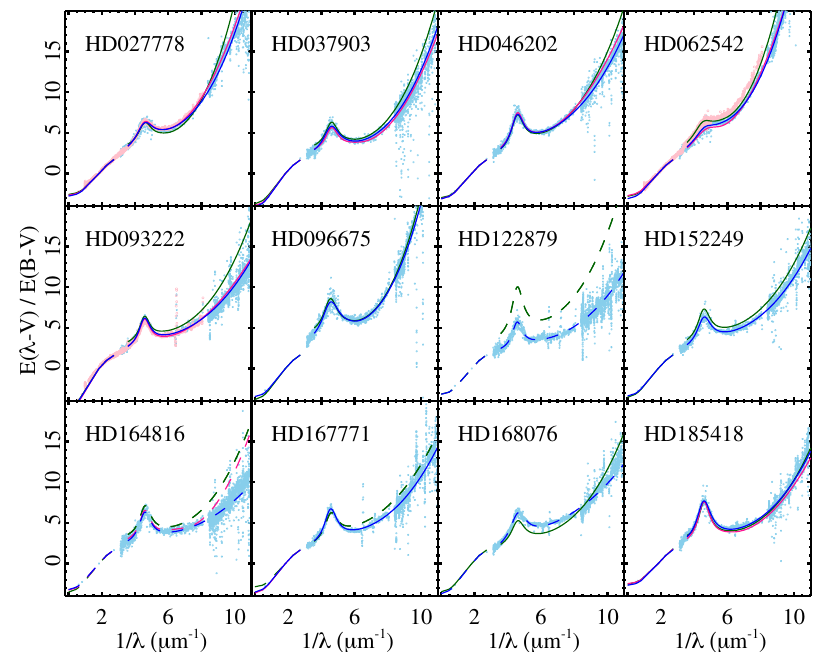}
 \end{center}   
   \caption{\label{reddG.pdf} { {Comparison of reddening curves derived
     for the same sightline. The symbols represent the reddening data
     from \citet{G09} shown as blue dots and by \citet{F19} in salmon,
     respectively. Reddening curves derived from $UV$ spline fits
     (Eq.~\ref{ext.eq}) and interpolated in $UBVJHK, -R_V$ are taken
     from \citet{G09}, shown as blue, by \citet{FM07} in red,
     and by \cite{V04} by green lines. Non peculiar stars of the
     high-quality sample listed in Table~\ref{ok.tab} are shown as
     full lines with the others as dashed lines.}}}
\end{figure*}

\begin{figure*} [h!tb]
  \begin{center}
    \includegraphics[width=18cm,clip=true, trim=4cm 6.4cm 3.5cm 6.9cm]{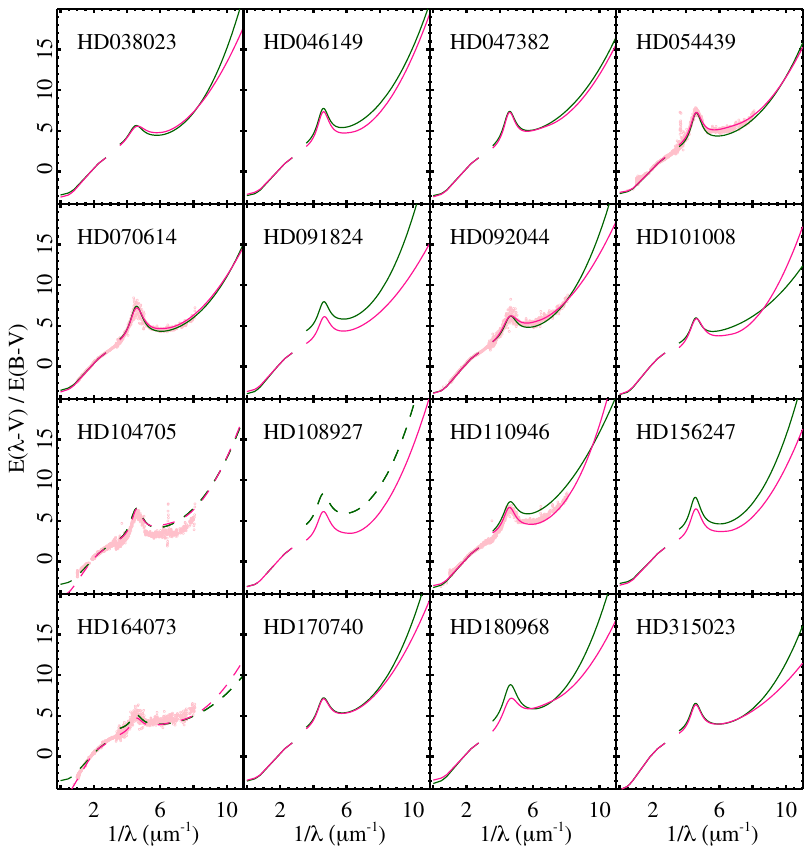}
  \end{center}
  \caption{\label{reddFV.pdf} {{Comparison of reddening curves
        derived for the same sightline. Notation as in
        Fig.~\ref{reddG.pdf}: \cite{V04} in green, \citet{FM07} in
        red, and \citet{F19} in salmon.}}}
\end{figure*}


\section{The high quality sample of reddening curves in the Milky Way \label{MWredd.sec}}

\subsection{The sample \label{ok.sec}}

Our high-quality sample contains only reddening curves that
respect the following six confidence criteria derived in
Sect.~\ref{scrut.sec}:

\smallskip
\noindent
 \/ a) The IUE spectra include only a single star that dominates the
 observed spectrum.

 \smallskip
 \noindent
 \/ b) The GAIA parallaxes $\pi_{\rm {DR3}}$ and  $\pi_{\rm {DR2}}$ 
  are consistent within their 3$\sigma$ error estimates.

 \smallskip
 \noindent
 \/ c) The GAIA photometric variability between DR3 and DR2 is within $-19
\simless \Delta G \simless 47$\,(mmag).

\smallskip
\noindent
 \/ d) The variability of the star observed from the ground and by
 Hipparcos in the $V$-band is within $-100 \simless \Delta V
 \simless 88$\,(mmag) and in the $B-V$ colour within $-55 \simless
 \Delta_{BV} \simless 112$\,(mmag), respectively.

 \smallskip
 \noindent
 \/ e) SpT and LC are derived as in Sect.~\ref{spclass.sec}
 at high confidence ($\chi^2 < 2$).

 \smallskip
 \noindent
 \/ f) The stellar classification of program and comparison star
    agrees within one subtype.

 With these confidence criteria, the high-quality sample { {is
     given in Table~\ref{ok.tab}. It}} comprises { {80}} reddening
 curves for { {53}} sightlines { {of which 35 are the rare cases
     of single-cloud dominated sightlines. Six reddening curves show a
     peculiar jump between 0.28$\mu$m and the $U$-band,
     i.e. $E$(0.28$\mu$m -$U)/E(B-V) > 2.2$\,mag.}} We give preference
 to reddening curves by \cite{G09} as they include besides IUE also
 FUSE data. We also prefer reddening curves by \cite{FM07} over \cite{V04}
 because comparison star observations that were used in the latter
 work are not needed in the former derivation procedure. Our choice of
 the reference reddening curve $E^{\rm{Ref}}$ for each sightline is
 given in column~3; when available, other accepted reddening curves
 $E^{\rm{i}}$ with $i \in \{\rm{FM07, V04}\}$ are listed in column~4.
 None of the stars of Table~\ref{ok.tab} is associated with reflection
 nebulosity within 5$''$ \citep{M03} as otherwise, the curves would
 not represent the reddening in the translucient cloud.

All reddening curves of both the high-quality sample
(Table~\ref{ok.tab}) and those of the stars with uncertainties in the
reddening curve are shown in Fig.~\ref{reddG.pdf} and
Fig.~\ref{reddFV.pdf}, respectively. A nearly perfect match for stars
with accepted curves by \citet{G09}, \cite{FM07}, and \cite{F19} are
visible in Fig.~\ref{reddG.pdf} and for about half of the stars with
high-quality reddening curves by \cite{FM07} and \cite{V04} in
Fig.~\ref{reddFV.pdf}. We note that the consistency between various
reddening curves of a star cannot be taken as full proof of the
correctness of the derivation. In the far UV a steeper rise in the
reddening is present for six curves derived by \cite{V04} and for
three stars by \cite{FM07} when compared to any other of the available
reddening curves for these stars. In the near IR differences in the
reddening between \cite{V04} and \cite{FM07} are visible for three
sightlines (Fig.~\ref{reddFV.pdf}).

\subsection{Intrinsic errors in the high-quality sample \label{systematics.sec}}

The intrinsic error of the reddening curves in the high-quality sample
(Sect.~\ref{ok.sec}) was estimated by measuring the variance between
individual reddening curves towards the same star.  For each star of
Table~\ref{ok.tab},{ { which is not flaged as peculiar}}, the ratio
$E_{\lambda}^{\rm i}/E_{\lambda}^{\rm {Ref}}$ of the two reddening
curves was computed, where the reference for the reddening curve
$E_{\lambda}^{\rm {Ref}}$ is given in column ~3 and the reference for
the reddening curve $E^{\rm{i}}$ with $i \in \{\rm{FM07, V04}\}$ in
column ~4 of Table~\ref{ok.tab}.  For example, for HD~027778 the ratio
of the reddening curves $E_{\lambda}^{\rm{FM07}}/E_{\lambda}^{\rm
  {G09}}$ were determined, for HD~030470 there exists no such ratio,
and for HD~037903 there are even two ratios available --
$E_{\lambda}^{\rm{FM07}}/E_{\lambda}^{\rm {G09}}$ and
$E_{\lambda}^{\rm{V04}}/E_{\lambda}^{\rm {G09}}$. In total there are
27 $(E^{\rm i}/ E^{\rm {Ref}})$ ratios of reddening curves that all
pass the confidence criteria of Sect.~\ref{ok.sec}.

\noindent
The mean and $1\,\sigma$ error of these $(E^{\rm i}/ E^{\rm {Ref}})$
ratios are computed by omitting data at $1/\lambda \, \simgreat
7.5\,\mu{\rm m}^{-1}$ when not observed by FUSE. In the $V$-band, to
which the curves are normalized, the scatter of these ratios is
$\sigma(E^{\rm i}/E^{\rm {Ref}}) = 0$ by definition and remains
naturally small for wavelengths close to it. In the IUE range
$\sigma(E^{\rm i}/E^{\rm {Ref}}) \sim 10$\,\% and grows to $\sim
15$\,\% in the far UV at $x \sim 11\,\mu$m$^{-1}$. Similarly
$\sigma(E^{\rm i}/E^{\rm {Ref}}) \simless 7$\,\% for $\lambda \sim
1\,\mu$m and increases to 11\% at longer wavelengths.  The typical
error in the high-quality sample when averaged over wavelengths is
$\sigma(E^{\rm i}/E^{\rm {Ref}}) \sim 9$\,\% and stays below 16\,\%.
Repeating the same exercise for the {{58}} stars with uncertain
reddening curves gives { {23}} such ratios and shows a larger
scatter of $\sigma(E^{\rm i}/E^{\rm {Ref}}) \sim 15$\,\%.  The ratios
of uncertain reddening curves of the same star varies by up to
$39\,\%$, which is about a factor of two larger spread than for the
high-quality sample.

\subsection{Uncertainties in $R_V$ \label{Rv.sec}}

\begin{figure} [h!tb]
\begin{center}
\includegraphics[width=8.5cm,clip=true,trim=4.cm 6.4cm 4.cm 8.3cm]{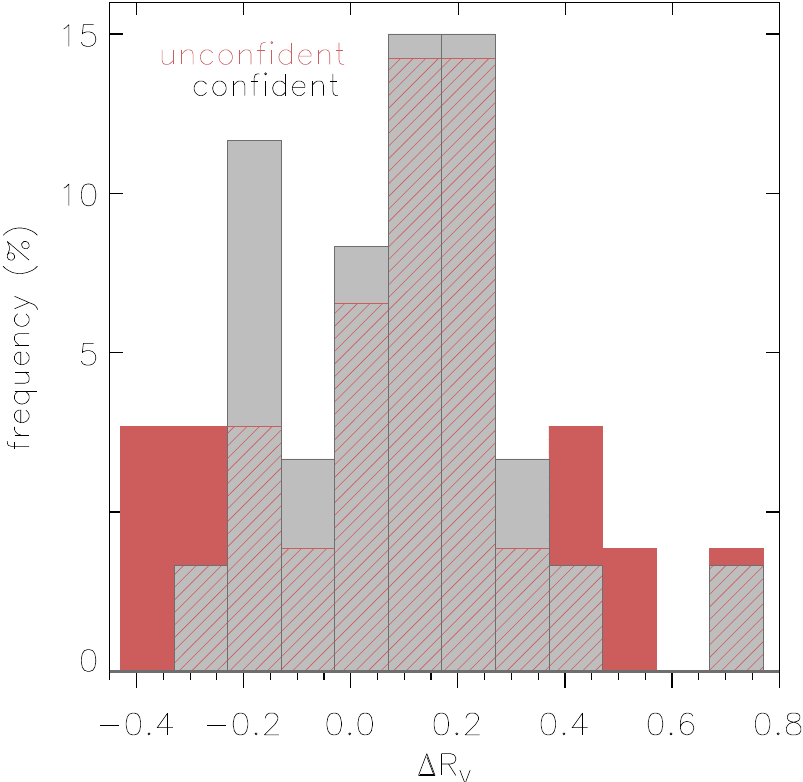}
\end{center}
\caption{\label{Rvar.pdf} Systematic differences in estimates of the
  total-to-selective extinction. The histograms show $\Delta R_{\rm V}
  = E^{\rm{ref}}(\infty) - E^{\rm i}(\infty)$ of individual stars in
  the sample with uncertain (red) and confident reddening curves
  (grey).}
\end{figure}

The shape of the extinction curves depends on the total-to-selective
extinction $R_V$ \citep{Cardelli89,Zagury}. { {This parameter is
    estimated by extrapolating a derived reddening, e.g. in the near or
    mid infrared, to infinite wavelengths; thus $R_V$ is not an
    observable. It successfully describes averaged properties with
    clear deviations in the reddening for specific sightlines
    \citep{Gordon23}.}}  The systematic uncertainties of $R_V = -
E(\infty)$ (Eq.~\ref{tau2k.eq}) are hence of interest.  The
differences $\Delta R_V = E^{\rm{ref}}(\infty) - E^{\rm i}(\infty)$ of
the available estimates of $R_V$ for the same star are computed. They
provide an estimate of the systematic errors and are shown in
Fig.~\ref{Rvar.pdf} as histograms for the sample of uncertain and
confident reddening curves. The references to the reddening curves
$E^{\rm{ref}}$ and $E^{\rm i}$ are listed for the high-quality sample
in colums~3 and 4 of Table~\ref{ok.tab}. For the unsecure sample we
use -- whenever available -- the estimate of $E^{\rm{ref}}(\infty)$ by
\cite{G09} and otherwise \cite{FM07}.  Three stars HD~104705,
HD~164073 and HD~147933 deviate with $\|\Delta R_V \| > 1.3$ and are
therefore omitted.  For example, HD~164073 has $R_V = 2.96$
\citep{V04} or $R_V = 5.18$ \citep{FM07}.  The distributions in
$\Delta R_V$ appear similar in both samples and are flatter than
Gaussian, which indicates that systematic errors dominate.  The
high-quality sample has a peak-to-peak scatter, mean, and 1$\sigma$
error of $-0.33 \, \simless \, \Delta R_V = 0.09 \pm 0.21 \, \simless
\, 0.67$. The reference reddening curves of the high-quality sample
vary between $2.01 \, \simless \, R_V = 3.08 \pm 0.44 \, \simless \,
4.34$ (Sect.~\ref{MW.sec}). We find that the derivation of $R_V$ for
the same star by the various authors agrees to better than
$\sigma(\Delta R_V) / \rm{mean(R_V)} \sim 7$\%.

\begin{table}[!htb]
\scriptsize   
\begin{center} 
  \caption {Rejected sightlines (58) with uncertain reddening
    curves. \label{reject.tab} }
\begin{tabular}{l c | l c}
\hline\hline
Star & flag & Star & flag \\
\hline
CPD$-$5926\, 00 & $\pi$		&
HD~024263  & M	                \\
HD~024912 & $\Delta G$ 		&
HD~030123  & $\Delta V$         \\ 
HD~037022 & M 	    		&
HD~037023  & M       	        \\ 
HD~037041 & $\Delta G$ 		&
HD~037130  & M 	                \\ 
HD~037367 & $\chi^2$ 		&
HD~046660  & O7~V 	        \\
HD~054306     & B3~V            &
HD~072648 &B5~Ib 	 	\\
HD~073882  & $\pi$ 	        & 
HD~075309 & $\chi^2$ 		\\
HD~091983  & $\Delta V$ 	&  
HD~093632 & M 			\\
HD~093843  & O4~Ib   	        & 
HD~094493 & $\chi^2$ 		\\
HD~096042 & $\Delta V$ 		&
HD~097484 & $\Delta V$ 		\\
HD~099872  & M 	 	        &  
HD~103779 &B2~Ib     		\\
HD~104705  & $\Delta R_V$       &
HD~122879   & B0~Ia             \\
HD~134591 & B4~V		&
HD~141318 & M 			\\
HD~143054  & $\pi$ 	        & 
HD~147165  & $\Delta G$ 	\\ 
HD~147701 & $\Delta_{BV}$ 	&
HD~147888   & $\Delta G$ 	\\  
HD~147889   & $\chi^2$ 		&
HD~147933   & M                 \\
HD~149757   & $\Delta G$ 	& 
HD~151804   & O8~Ia     	\\
HD~153919   & O4~Ib 		& 
HD~154445   & $\chi^2$	        \\
HD~162978   & O8~II 		&  
HD~163181   & $\Delta V$        \\
HD~164073   & $\Delta R_V$      &
HD~164536   & O7~V	        \\
HD~164816   & O8~V              &
HD~164906   & M 	        \\
HD~164947A  & M 	  	& 
HD~164947B  & M                 \\
HD~167838   & $\chi^2$ 		& 
HD~168941   & $\pi$             \\
HD~169454   & $\Delta_{BV}$ 	& 
HD~175156   & $\chi^2$ 	        \\
HD~185859   & B2~Ib 		& 
HD~198478   & $\pi$ 	        \\
HD~203532   & B5~III            & 
HD~204827   & M 	        \\
HD~210121   & $\chi^2$ 		& 
HD~294264   & $\Delta V$        \\
HD~315031    & $\Delta V$       &
HD~315033   & B3~V	  	\\
Herschel~36 & $\Delta G$        &
Walker~67   & $\chi^2$ 		\\
\hline
\end{tabular}
\end{center}
\scriptsize{ {\bf Notes:} Criteria leading to the rejection of a
  sightline are indicated by flags: M, when equally bright objects
  were in the IUE aperture, $\pi$, for variations in the GAIA
  parallax, $\Delta G$, $\Delta V$, and $\Delta_{BV}$ for photometric
  variability, $\Delta R_V$, for discrepant $R_V$ estimates, $\chi^2$,
  when the SpT of the reddened star cannot be estimated securely, and
  the SpT estimates of Sect.~\ref{spclass.sec} is given, when it
  deviates from that used in the reddening determination.}
\end{table}


\begin{table}[!htb]
\begin{center}
\caption {The high-quality Milky Way reddening curve
  sample. \label{ok.tab} } \scriptsize
\begin{tabular}{l l l l c}
\hline\hline
Name & SpL & $E^{\rm{Ref}}$ & $E^{\rm{i}}$ &  Clouds$^{\dagger}$\\
\hline
    HD~027778     & B3~V          & G09   & FM07      & M \\
    HD~030470     & B9~V          & FM07  &  -        & S \\
    HD~030492   & A0~III          & FM07  &  -        & S \\
    HD~037903   & B2~V            & G09   & FM07, V04 & M \\
    HD~038023     & B3~V          & FM07  & V04       & S \\
    HD~046149     & O8~V          & FM07  & V04       & M \\
    HD~046202     & O9~V          & G09   & FM07, V04 & M \\
    HD~046223     & O4~V          & V04   &  -        & S \\
    HD~047382    & B0~Ib          & FM07  & V04       & M \\
    HD~054439     & B1~V          & FM07  & V04       & S \\
    HD~062542     & B5~V          & G09   & FM07, V04 & S \\
    HD~070614     & B3~III        & FM07  & V04       & M \\
    HD~079186     & B3~Ia         & {{V04$^E$}}   & - & S  \\
    HD~089137     & B0~Ib         & {{V04$^E$}}   & - & S  \\
    HD~091824     & O7~V          & FM07  & {{V04$^E$}}   & M \\
    HD~092044    & B1~Ib          & FM07  & V04       & S \\
    HD~093205     & O4~V          & V04   &  -        & M \\
    HD~093222     & O7~III          & G09   & FM07, V04 & M \\
    HD~096675   & B7~V          & G09   & -         & S \\
    HD~101008   & B0~III          & FM07  & V04       & S \\
    HD~108927     & B5~V          & FM07  &  -        & S \\
    HD~110336    & B8~V           & FM07  &  -        & S \\
    HD~110715   & B9~IV           & FM07  &  -        & S \\
    HD~110946   & B2~III          & FM07  & V04       & S \\
    HD~112607   & B5~III          & FM07  &  -        & S \\
    HD~112954   & B9~III          & FM07  &  -        & S \\
    HD~129557     & B1~V          & V04   &  -        & S \\
    HD~146284    & B9~III         & FM07  &  -        & S \\
    HD~146285    & B9~IV          & FM07  &  -        & S \\
    HD~147196     & B8~V          & FM07  &  -        & S \\
    HD~148379    & B2~Ia          & V04   &  -        & M \\
    HD~148579    & B9~IV          & FM07  &  -        & S \\
    HD~148594   & B7~V            & FM07  &  -        & S \\
    HD~149038   & B0~Ib          & {{V04$^E$}}   & -         & M \\
    HD~152235    & B1~Ia          & V04   &  -        & M \\
    HD~152245   & B0.5~III        & V04   &  -        & M \\
    HD~152249    & O9~Ia          & G09   & V04       & S \\
    HD~156247     & B5~III        & FM07  &  {{V04$^E$}}        & S \\
    HD~164402    & B0~Ib          & V04   &  -        & S \\
    HD~167264    & B0~Ib          & V04   &  -        & S \\
    HD~167771   & O7~III          & G09   & FM07      & S \\
    HD~168076   & O4~III/O5~V     & {{V04}}   &  -        & M \\
    HD~170634     & B8~V          & FM07  &  -        & M \\
    HD~170740     & B2~V          & FM07  & V04       & M \\
    HD~180968   & B0.5~III        & FM07  &  {{V04$^E$}}       & S \\
    HD~185418     & B0.5~V        & G09   & FM07, V04 & M \\
    HD~287150   & A1~III          & FM07  &  -        & S \\
    HD~294304     & B6~V          & FM07  &  -        & S \\
    HD~303308   & O6~V            & V04   &  -        & M \\
    HD~315021     & B0~V          & FM07  & -         & S \\
    HD~315023   & B2~V            & FM07  & V04       & S \\
    HD~315024     & B1~V          & FM07  &  -        & S \\
    HD~315032     & B1~V          & FM07  & V04       & S \\
\hline 
\end{tabular}
\end{center}
\scriptsize {{\bf Notes.} The sample includes {{80 reddening curves
      for 53 sightlines. Six curves show a peculiar jump in $0.28
      \mu\rm{m}$ and the $U$-band and are maked by $^E$ .}}  Column
  SpL refers to SpT and LC estimates of Sect.~\ref{spclass.sec}.
  Reddening curves by G09 \citep{G09}, FM07 \citep{FM07}, and V04
  \citep{V04} that pass the six confidence criteria of
  Sect.~\ref{scrut.sec} are given in columns $E^{\rm{Ref}}$ as our
  first choice and $E^{\rm{i}}$ when available. $^{\dagger}$
  Classification as single-cloud (S) or multiple-clouds (M) sightlines
  is taken from \cite{S20}.}
\end{table}



\subsection{Reddening in the Milky Way \label{MW.sec}}

Galactic extinction studies derive the typical wavelength dependence
of interstellar reddening. Such curves are commonly used, in
particular in extra-galactic research, as the standard for dereddening
the observed flux of objects for which there is no specific knowledge
about the dust. Large variations from cloud-to-cloud of the grain
characteristics and the physical parameters of the dust are reported
by \citet{S18}. Although averaging a sufficiently large number of
clouds and sightlines leads to similar mean parameters, they likely do
not reflect the true nature of the dust. The degree to which mean
Galactic extinction curves can be taken as typical shall always be put
in question.


\begin{figure} [h!tb]
\begin{center}
\includegraphics[width=9.5cm,clip=true,trim=4.4cm 6.5cm 3.2cm 7.2cm]{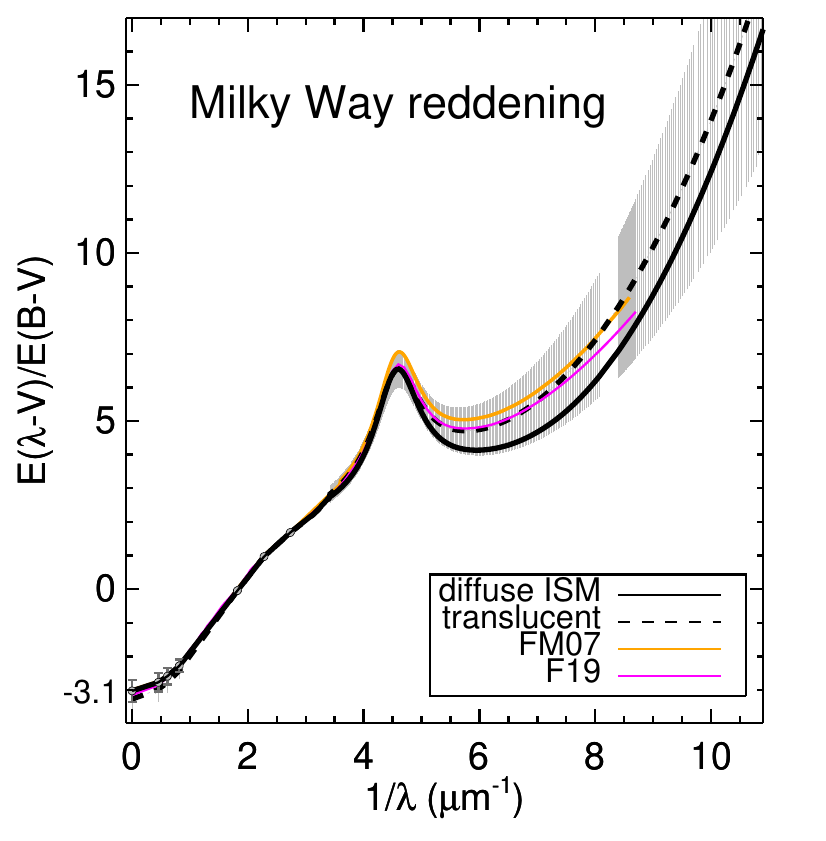}
\end{center}
\caption{\label{pl_meanISM.pdf} The Milky Way
  reddening derived from the high-quality sample (Table~\ref{ok.tab})
  for translucent clouds (dashed with 1$\sigma$ error bars) and for
  single-cloud sightlines of the diffuse ISM. The average Galactic
  reddening derived by \cite{FM07} and \cite{F19} are shown for comparison.}
\end{figure}


The shape of the average Galactic extinction for a sample of 243 stars
with $2.4 \leq R_V \leq 3.6$ at $1/\lambda \leq 8.6\,\mu{\rm m}^{-1}$
has been presented by \citet{FM07}. Their mean curve has $R_V = 3.1$
as derived earlier \citep{FM90, V04} and in recent studies
\citep{F19}; \citet{Wang19} find a similar value of $R_ V =
3.16$. Here we compute average Milky Way reddening curves using the
high-quality sample (Table~\ref{ok.tab}, col.~3). As in the previous
sections a $3\sigma$ rejection criterium is applied, so that stars
have $A_V \simless 2.4$ and $R_V \simless 4.4$. HD~093222 and
HD~168076 violate this criterium and are not included in the
analysis. We distinguish sightlines of translucent clouds at $1 <
\tau_V \simless 2.2$ and single-cloud sightlines of the diffuse ISM at
$\tau_V \simless 1$.  The average curve of the 33 translucent clouds
in our sample has a peak-to-peak scatter, mean, and 1$\sigma$ error of
$2.7 \, \simless \, R_V = 3.3 \pm 0.4 \, \simless \, 4.1$ at median of
$R_V = 3.16$. That of the 15 single-cloud diffuse ISM sightlines has
$2.3 \, \simless \, R_V = 3.0 \pm 0.3 \, \simless \, 3.6$ at median of
$R_V = 3.1$, respectively. These curves are shown in
Fig.~\ref{pl_meanISM.pdf} with the average Galactic reddening by
\citet{FM07} and \citet{F19}. Driven by the same $R_V$ of these curves
a nearly perfect match from optical to longer wavelengths is found;
for translucent clouds this incldues the IUE range and for the diffuse
ISM the reddening is smaller at $\lambda < 0.2\mu\rm{m}$. Noticeable
is the large diversity of the reddening curves
(Fig.~\ref{redd_all.pdf}) and agreement of the various mean Milky Way
reddening curvesl within $\sigma(R_{\rm V}) = 0.4$.

\section{Summary \label{summary.sec}}

Individual clouds in the ISM can be drastically different from each
other. Therefore, the framework of single-cloud sightlines was
introduced as they provide an unambiguous view of physical relations
between dust properties and observables such as extinction and
polarisation \citep{S18, S20}. Reddening curves allow to
derive fundamental characteristics of dust. However, before
extracting dust properties from a physical model one must ensure 
that the observational basis is solid and the assumptions in the 
derivation of the reddening are met to an acceptable level.

We discussed the current database of available reddening curves. The
initial sample of 895 reddening curves towards 568 OB stars, which
cover the spectral range from the near IR to the Lyman limit was
merged with 186 OB stars with high-resolution UVES spectra 
and with polarisation spectra towards 215 OB stars from
the Large Interstellar Polarisation survey. This yields a sample of
111 sightlines for which the reddening curves were scrutinized against
systematic errors by the following means:

\smallskip
\noindent
\/ a) Whenever IUE/FUSE spectra were identified as a composite of
multiple sources the corresponding reddening curves were rejected.
Stars with assigned binary information did not provide a direct link
or a striking feature in the appearance of the reddening curves for
qualifying a rejection.

\smallskip
\noindent
 \/ b) The stellar classification of the stars was derived at high
 confidence. Objects whose spectral types do not match within one
 subtype the comparison star were removed from the sample.

\smallskip
\noindent
 \/ c) Stars with detected variability between $1950 - 2017$ from
 ground-based and Hipparcos $V$-band photometry and $B-V$ colurs of
 about $0.1$\,mag were excluded. The same holds for GAIA $G$-band
 photometric variations of more than 47\,mmag.

\smallskip
\noindent
 \/ d) Reddening curves of stars showing inconsistencies in the GAIA
 parallaxes between DR2 and DR3 were declared as spurious.

 \smallskip
\noindent
 \/ e) The reddening curves with different estimates of $R_V$ of the
 same star exceeding 50\% were also rejected. \\

In total, we find { {53}} stars with one or more reddening curves
passing the rejection criteria of which {{35}} are the rare cases
of single-cloud sightlines. This provides the highest quality Milky
Way reddening curve sample available today. The average Milky Way
reddening curve is determined for translucent clouds and the diffuse
ISM to be $R_V = 3.1 \pm 0.4$, confirming earlier estimates. The
high-quality reddening curve sample together with polarisation
properties will be subject to dust modelling in a future paper in this
series of the Dark Dust project.

\begin{acknowledgements}
JK acknowledge the financial support of the Polish National Science
Centre, Poland (2017/25/B/ST9/01524) for the period 2018 - 2023. This
research has made use of the services of the ESO Science Archive
Facility and the SIMBAD database operated at CDS, Strasbourg, France
(Wanger et al. 2000) and partially based on observations collected at
the European Southern Observatory under ESO programmes
(Sect.~\ref{Prog.id}).
\end{acknowledgements}

\appendix

\section{Comments on indiviudal stars \label{comm.ap}}

Reddening curves with a miss-match of more than one type between our
estimates (Sect.~\ref{spclass.sec}) and the SpT used in the derivation
are rejected by the following reason:

\smallskip
\noindent HD~054306: we classify it as B3V whereas \cite{FM07} finds
it at 22,409\,K so earlier than B1.5~I and \citet{V04} used a B1~V
comparison star { {(HD~031726)}}.

\smallskip
\noindent HD~072648: we classify it as B5~Ib whereas \cite{FM07} finds
it at 21,035\,K so earlier than B1.5~I and \citet{V04} used a B1.5~III
comparison star.

 \smallskip \noindent { {HD~091983: we classify this photometric
     variable star (Fig.~\ref{sel_EBV.pdf}) as O9~V, \cite{Sota14} as
     O9~IV, and \cite{Cardelli92} finds it as O9.5~Ib.}}

\smallskip
\noindent HD~096675: we classify it as B7~V, \citet{G09}
as B6~IV and \citet{V04} used a B5~V comparison star.

\smallskip
\noindent HD~103779: we classify it as B2~Ib whereas \citet{G09, V04}
used a B0.5~Ib comparison star.

\smallskip
\noindent HD~134591: we classify it as B8~III whereas \citet{V04} used
a B4~V comparison star.

\smallskip
\noindent HD~151804, HD~153919, HD~162978: we classify these stars as
O~8Ia, O4~Ib, O8~II and \cite{Sota14} as O8~Iab, O6~V, O8~III; whereas
\citet{V04} used a O~9.5Ia comparison star.

\smallskip
\noindent HD~164536: we classify it, in agreement to \cite{Sota14}, as
O7~V whereas \cite{FM07} finds it at 33,500\,K, close to O9~V
  \citep{Theo91}.

\smallskip
\noindent HD~164816: we classify it as O8~V, \cite{Sota14} as
O9.5~V+B0~V binary, \citet{V04} as B0~V, and \cite{FM07} finds it at
31,427 K, close to O9.5~V \citep{Theo91}, \citet{G09} used HD~097471 a
{ {B0~V}} comparison star.

\smallskip
\noindent { {HD~168076: we classify it as O5~V at $\chi^2=0.31$ and
    as O4~III at $\chi^2=0.17$ in agreement to \cite{Sota14}.}}

\smallskip
\noindent HD~185859: we classify it as B2~Ib whereas \citet{V04} used
a B0.5~Ib comparison star.

\smallskip
\noindent HD~203532: we classify it as B5~III, \cite{FM07}
finds it at 17,785\,K so earlier than that \citep{Theo91} and
  \citet{V04} used a B3~IV comparison star.

\smallskip
\noindent HD~315033: we classify it as B3~V whereas \cite{FM07} finds
it at 25,609\,K so earlier than B1.5~V and \citet{V04} used a B1.5~III
comparison star.

\section{ESO Programmes \label{Prog.id}}

Observations collected at the European Southern Observatory are based
under ESO programmes: {\scriptsize{060.A-9036(A), 067.C-0281(A),
071.C-0367(A), 072.D-0196(A), 073.C-0337(A), 073.C-0337(A),
073.D-0609(A), 074.D-0300(A), 075.D-0061(A), 075.D-0369(A),
075.D-0369(A), 076.C-0164(A), 076.C-0431(A), 076.C-0431(B),
079.D-0564(A), 081.C-0475(A), 081.D-2008(A), 082.C-0566(A),
083.D-0589(A), 083.D-0589(A), 086.D-0997(B), 091.D-0221(A),
092.C-0019(A), 102.C-0040(B), 102.C-0699(A), 187.D-0917(A),
194.C-0833(A), 194.C-0833(B), 194.C-0833(D), 194.C-0833(E),
194.C-0833(F), 194.C-0833(H), 
{ { 0102.C-0040(B), 072.A-0100(A), 072.B-0123(D), 072.B-0218(A),
     072.C-0488(E), 076.B-0055(A), 077.C-0547(A), 078.D-0245(C),
     079.A-9008(A), 079.C-0170(A), 083.A-0733(A), 088.A-9003(A),
     089.D-0975(A), 091.C-0713(A), 091.D-0221(A), 092.C-0218(A),
     096.D-0008(A), 097.D-0150(A), 106.20WN.001, 1102.A-0852(C),
     165.N-0276(A), 194.C-0833(A), 194.C-0833(B), 194.C-0833(C),
     266.D-5655(A), 65.H-0375(A), 65.N-0378(A), 65.N-0577(B),
     70.D-0191(A)}.
}}

\bibliographystyle{aa}
\bibliography{References}

\end{document}